\documentclass[numsec,webpdf,modern,large]{oup-authoring-template}

\usepackage{natbib}

\setcitestyle{authoryear,round,aysep={}}

\graphicspath{{Fig/}}

\theoremstyle{thmstyleone}%
\theoremstyle{thmstyletwo}%
\theoremstyle{thmstylethree}%

\journaltitle{}
\copyrightyear{2026}

\usepackage{xr}

\newenvironment{namedlist}[1][0.4cm]{%
  \begin{list}{}%
  {%
    \setlength{\labelwidth}{#1}%
    \setlength{\leftmargin}{#1}%
    \setlength{\labelsep}{0.5em}%
    \setlength{\itemindent}{0pt}%
  }%
}{%
  \end{list}%
}

\usepackage[caption=false]{subfig}
\usepackage{graphicx}
\usepackage[export]{adjustbox}
\usepackage{relsize}
\usepackage{grfext}
\usepackage{amsmath}

\newcommand{\myBigFig}[3][0.7]{
\begin{figure*}[tbp]
\centering
\includegraphics[max width=#1\textwidth, max height=0.9\textheight, keepaspectratio]{#2}
\caption{#3}
\label{fig:#2}
\end{figure*}
}

\newcommand{\figRef}[1]{Fig.\@ \ref{fig:#1}}

\newcommand{\pseudocodeKeyword}[0]{Algorithm}

\newcommand{\supPseudocodeKeyword}[0]{Supplementary Algorithm}

\floatstyle{ruled}
\newfloat{codeFloat}{tbp}{cod}
\floatname{codeFloat}{\pseudocodeKeyword{}}
\newfloat{supCodeFloat}{tbp}{cod}
\floatname{supCodeFloat}{\supPseudocodeKeyword{}}

\newcommand{\paperTitle}[0]{Genetic algorithms for multi-omic feature selection: a comparative study in cancer survival analysis}
\newcommand{\shortTitle}[0]{Genetic algorithms for multi-omic feature selection}

\newcommand{\algorithmStart}[2]{
	\begin{algorithm}[htbp]
	\caption{#1}\label{code:#2}
	\begin{algorithmic}[1]}

\newcommand{\algorithmEnd}{
	\end{algorithmic}
	\end{algorithm}}
	
\newcommand{\supAlgorithmStart}[2]{
	\begin{algorithm}[H]
	\caption{#1}\label{supCode:#2}
	\begin{algorithmic}[1]}

\newcommand{\supAlgorithmEnd}{
	\end{algorithmic}
	\end{algorithm}}

\begin{document}

\DOI{DOI HERE}
\pubyear{2026}
\access{Advance Access Publication Date: Day Month 2026}
\appnotes{Paper}

\firstpage{1}


\title[\shortTitle{}]{\paperTitle{}}

\author[1,$\ast$]{Luca Cattelani\ORCID{0000-0003-4852-2310}}
\author[1,$\ast$]{Vittorio Fortino\ORCID{0000-0001-8693-5285}}

\authormark{L.\,Cattelani and V.\,Fortino}

\address[1]{\orgdiv{Institute of Biomedicine, School of Medicine}, 
\orgname{University of Eastern Finland}, 
\orgaddress{\street{Yliopistonranta 8}, \postcode{70210}, \country{Finland}}}

\corresp[$\ast$]{Correspondence: \href{mailto:luca.cattelani@uef.fi}{luca.cattelani@uef.fi}; 
\href{mailto:vittorio.fortino@uef.fi}{vittorio.fortino@uef.fi}}

\received{Date}{0}{2026}
\revised{Date}{0}{2026}
\accepted{Date}{0}{2026}

\abstract{
\textbf{Motivation:} Abstract should not exceed 250 words.\\
\textbf{Results:} .\\
\textbf{Availability and implementation:} .\\
\textbf{Contact:} \href{vittorio.fortino@uef.fi}{vittorio.fortino@uef.fi}\\
\textbf{Supplementary information:} Supplementary data are available.
}

\abstract{
Multi-omic datasets offer opportunities for improved biomarker discovery in cancer research, but their high dimensionality and limited sample sizes make identifying compact and effective biomarker panels challenging. Feature selection in large-scale omics can be efficiently addressed by combining machine learning with genetic algorithms, which naturally support multi-objective optimization of predictive accuracy and biomarker set size. However, genetic algorithms remain relatively underexplored for multi-omic feature selection, where most approaches concatenate all layers into a single feature space. To address this limitation, we introduce Sweeping*, a multi-view, multi-objective algorithm alternating between single- and multi-view optimization. It employs a nested single-view multi-objective optimizer, and for this study we use the genetic algorithm NSGA3-CHS. It first identifies informative biomarkers within each layer, then jointly evaluates cross-layer interactions; these multi-omic solutions guide the next single-view search. Through repeated sweeps, the algorithm progressively identifies compact biomarker panels capturing cross-modal complementary signals. We benchmark five Sweeping* strategies, including hierarchical and concatenation-based variants, using survival prediction on three TCGA cohorts. Each strategy jointly optimizes predictive accuracy and set size, measured via the concordance index and root-leanness. Overall performance and estimation error are assessed through cross hypervolume and Pareto delta under 5-fold cross-validation. Our results show that Sweeping* can improve the accuracy-complexity trade-off when sufficient survival signal is present and that integrating omic layers can enhance survival prediction beyond clinical-only models, although benefits remain cohort-dependent.
}

\maketitle

\thispagestyle{empty}

\section{Introduction}

Feature selection is central to biomarker discovery from large-scale omics data,
where machine learning (ML) models can capture complex associations
among molecular features and clinical outcomes beyond simple univariate filters \citep{Li2018}.
Omics datasets typically exhibit extreme dimensionality with limited samples,
making feature selection indispensable for both performance and interpretability.

Heuristic search approaches, particularly genetic algorithms (GAs),
have shown the ability to uncover small yet informative biomarker panels
\citep{Cattelani2022,Cattelani2024BiB,Cattelani2024JBI,Cattelani2024IEEE}.
Identifying such panels is especially important in multi-omics studies,
where integrating complementary molecular layers can improve the discovery of biomarkers
associated with clinically relevant outcomes, such as drug response and patient survival \citep{misra2019}.
However, extracting robust predictive signals across multiple omic modalities
has historically proven challenging due to the high dimensionality,
heterogeneity, and limited sample sizes typical of these datasets \citep{herrmann2021,Li2022,babu2023}.
Consequently, the simultaneous selection of ML-derived biomarkers together with the most relevant omics layers,
while aiming to identify minimal and interpretable multi-omics biomarker panels,
remains relatively underexplored \citep{Reel2021,Morabito2025}.

A common strategy for integrating multi‑omic data is to concatenate features
from different modalities into a single unified representation.
While straightforward, this approach substantially expands the dimensionality
of the search space and often leads the optimization process to select an excessive number of features,
typically dominated by the largest omic layer.
As a result, the derived biomarker panels may obscure the relative contribution of each modality
and fail to achieve the desired level of parsimony.
A more structured alternative is to organize the optimization process hierarchically.
In such a design, feature selection is first conducted independently within each omic layer
to identify informative candidates, followed by a multi‑omic refinement step in which cross‑modal
interactions are evaluated. Iteratively alternating between single‑omic and multi‑omic stages
can help guide the search toward more balanced and compact biomarker panels,
while better leveraging complementary information across data types.
However, a hierarchical strategy may also risk assigning disproportionately similar
representations to all modalities, even in scenarios where their predictive importance is strongly uneven.

We introduce Sweeping*, a novel, configurable multi-view (MV), multi-objective (MO)
optimization algorithm that alternates single‑view (SV) optimizers
with a MV master step that jointly refines solutions across all views.
SV phases explore candidate biomarkers independently within each omics layer,
while the MV step evaluates and refines their joint performance across modalities.
The MV results influence the populations used in subsequent SV runs,
favoring features that show stronger complementarity with other omics layers
and guiding the search toward more effective multi-omics biomarker panels.
SV and MV optimizers are iteratively alternated (here referred to as sweeps),
allowing the biomarker search to progressively focus on features that perform well
within their own modality while also contributing effectively to integrated multi-omics panels (Supplementary Section 1.1).

For benchmarking purposes, we consider survival prediction in cancer patient datasets from
The Cancer Genome Atlas (TCGA) \citep{hutter2018cancer},
using clinical variables together with two molecular data layers, mRNA expression and miRNA profiles.
The objective is to assess whether molecular features provide
additional prognostic information beyond established clinical variables,
a multi-omic biomarker discovery task where previous attempts have produced
only modest or inconsistent improvements \citep{lopez2019,Li2024}.  

Within the Sweeping* framework, the SV optimizer used in this study is NSGA3-CHS,
an MO GA designed to improve the precision of Pareto-front approximations in biomarker discovery tasks \citep{Cattelani2024BiB}.
Building on NSGA-III \citep{Deb2014} and NSGA2-CHS \citep{Cattelani2022},
it enables efficient exploration of high-dimensional feature spaces.
The MO optimization targets two complementary goals:
maximizing the concordance index (C-index) for survival prediction and maximizing root-leanness,
a measure of feature-selection parsimony \citep{Cattelani2024BiB}.
Performance is assessed using cross hypervolume (CHV),
which summarizes the overall quality of the approximated Pareto front under cross-validation \citep{Cattelani2024IEEE},
and Pareto delta ($P_{\Delta}$), which quantifies estimation error and reflects solution robustness \citep{Cattelani2024BiB}.

Overall, sweeping strategies tend to improve performance when survival datasets
contain a sufficient number of events.
Our results also demonstrate that,
when combined with MO ML, models integrating clinical and omic data,
and in some cases omics-only models, can outperform models based solely on clinical variables.

\section{Materials and methods}\label{sec:methods}

\subsection{Overview of the Sweeping* framework}

Sweeping* is a MV, MO optimization algorithm
designed to integrate heterogeneous data layers (``views''), such
as clinical, mRNA, and miRNA features, while jointly optimizing
multiple objectives like predictive accuracy and biomarker parsimony.
Its defining property
is the use of alternating optimization phases: at every sweep,
Sweeping* performs (i) a SV optimization step,
where each view is optimized independently, followed by (ii) a
MV optimization step, where all views are jointly refined.
This SV$\rightarrow$MV alternation enables the framework to exploit
view-specific signals while capturing complementary information
across omics layers. A final optional MV tuning step further refines
the integrated solutions before returning the final
approximated Pareto-optimal set.

A ``view'' corresponds to one data modality, and a ``solution'' is a
candidate subset of features used by the prediction model. A
population of solutions is a collection of such candidates, and a
MV solution is constructed by merging the view-specific
subsets produced during the SV steps. This representation enables
the algorithm to naturally include or exclude entire data views,
depending on their utility for improving the accuracy–complexity
trade-off.

A high-level schematic of the Sweeping* workflow is provided in Supplementary Fig. S1.
An UML class diagram summarizing the main software components is in Supplementary Fig. S2.
Implementation details, software
interfaces, and algorithmic pseudocode are provided in
Supplementary Section 1.1.

\subsection{Sweeping* optimization cycle}

Each execution of Sweeping* consists of a series of \emph{sweeps}, where
each sweep contains one SV and one MV phase:
\begin{namedlist}
  \item[\textbf{Single-view optimization (SV step)}]  
    Each layer is optimized independently using a
    MO optimization algorithm. This step refines feature
    subsets within each view without influence from other layers.
  \item[\textbf{Multi-view optimization (MV step)}]  
    The view-specific solutions are merged into integrated
    MV candidates. These merged solutions are further refined
    to exploit cross-view complementarities that cannot be captured
    by SV optimization alone.
\end{namedlist}

Sweeping* repeats this SV$\rightarrow$MV sequence for a user-defined number
of sweeps, enabling progressive refinement of both individual-layer
and integrated MV components. After the final sweep, an
optional MV \emph{tuning} pass applies an additional MO
optimizer to further stabilize and improve the approximated Pareto set.

The SV and MV optimizers used in each sweep, the number of
generations allocated to each phase, and whether tuning is applied
are configured by the user that specifies a sweeping strategy (Supplementary Section 1.1.1).
The main pseudocode definition of Sweeping* is in Supplementary Algorithm 1.

\subsection{Sweeping* configurations evaluated}

We evaluated five configurations of Sweeping* that differ in how
sweeps are structured and whether a tuning phase is applied.
All configurations use the recently proposed MO GA NSGA3-CHS \citep{Cattelani2024BiB}
in the internal optimization phases
and, for a fair comparison, always allocate a total of 300 generations,
divided across SV and MV phases according to the sweeping strategy
(Supplementary Table S1).
This choices ensure that differences in
performance are attributable to MV integration rather than
to differences in the underlying optimization engine.
The configurations include:

\begin{namedlist}
    \item[\textbf{Concatenated}] Does not use sweeps and applies a direct MV optimization.
    \item[\textbf{Resampled Sweeping (Sw)}] Alternates SV optimizations with MV resamplings.
    The MV step optimizes feature sets that are the union of SV feature sets computed by the previous
    SV step.
    \item[\textbf{Concatenated Sweeping (CSw)}] The
    MV step starts from feature sets that are the union of SV feature sets but then, differently
    from the resampled approach, proceeds
    in a concatenated manner, operating at the level of the single features.
    \item[\textbf{Lean Concatenated Sweeping (LCSw)}] Similar to the CSw,
    but the MV steps instead of starting from feature sets that are the union of SV feature sets,
    start instead from subsets of them (features are discarded at random).
    \item[\textbf{Resampled Sweeping with Tuning (SwT)}] Similar to the Sw, but with a
    concatenated MV tuning at the end.
\end{namedlist}

A full algorithmic description is in Supplementary Section 1.1, with the considered alternatives for
the MV step in Supplementary Section 1.1.3 and Supplementary Section 1.1.4.

These variants allow us to investigate how different sweeping strategies
influence biomarker discovery
performance.

\subsection{Data and multi-objective formulation}

We benchmarked the Sweeping* variants on three cancer types from
The Cancer Genome Atlas (TCGA): kidney renal clear cell carcinoma
(KIRC), brain lower grade glioma (LGG), and sarcoma (SARC).
For each cohort we used three data layers---clinical variables,
mRNA expression, and miRNA expression (more details in
Supplementary Section 1.4, event accumulation patterns in Supplementary Fig. S3).

Each MO optimization seeks to identify feature subsets
that jointly maximize survival predictive accuracy and minimize
feature set size. Predictive accuracy was measured using the
C-index, computed using a Cox proportional
hazards model. Parsimony was quantified using the root-leanness
metric, which penalizes larger biomarker sets \citep{Cattelani2024BiB}.
The quality of
the resulting Pareto front approximations was assessed using two complementary
metrics: CHV as a measure of overall performance that takes cross-validation into account \citep{Cattelani2024IEEE},
and Pareto delta ($P_{\Delta}$), which
quantifies the overestimation: discrepancy between training and testing fronts \citep{Cattelani2024BiB}.

All evaluations used 5-fold cross-validation for model assessment
and nested validation for external performance estimation.

\section{Results}

\subsection{Overall performance and overestimation}

\begin{figure*}[ht!]
    \centering
    
    \subfloat[\label{fig:chv}]{
        \includegraphics[width=0.9\textwidth]{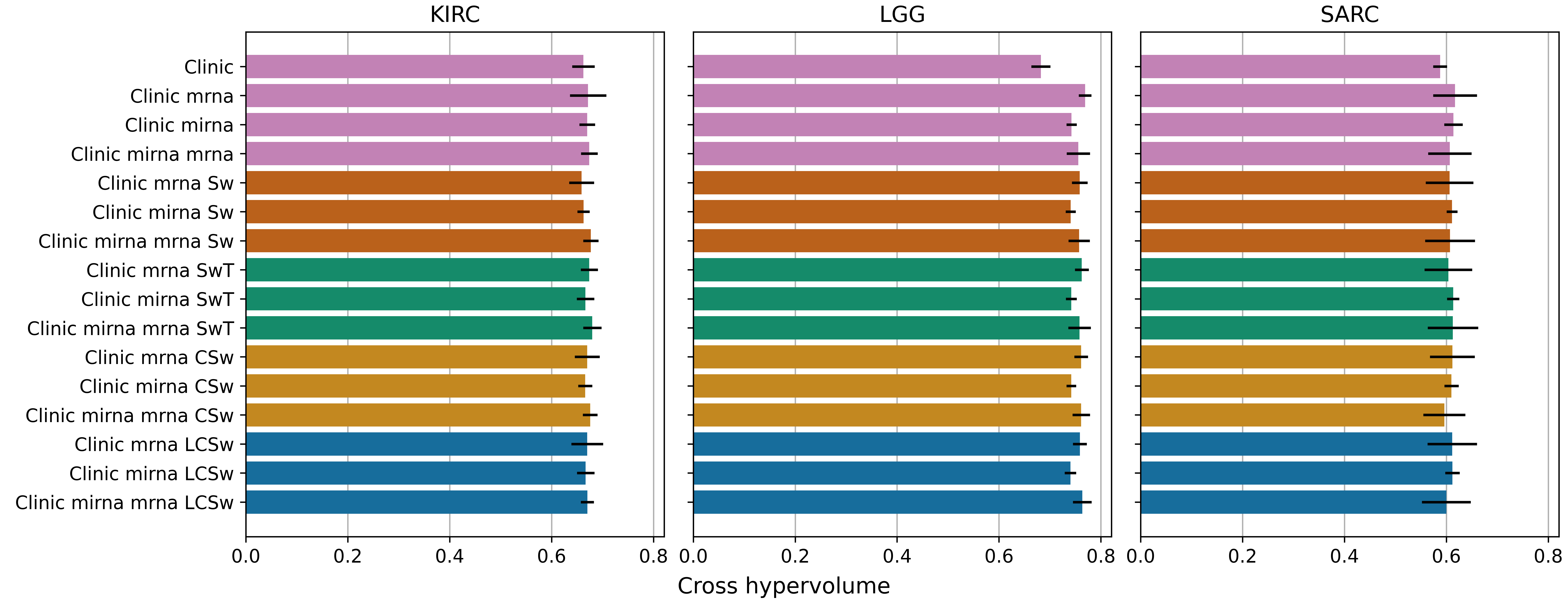}
    }\\[0.75em]
    \subfloat[\label{fig:pd}]{
        \includegraphics[width=0.9\textwidth]{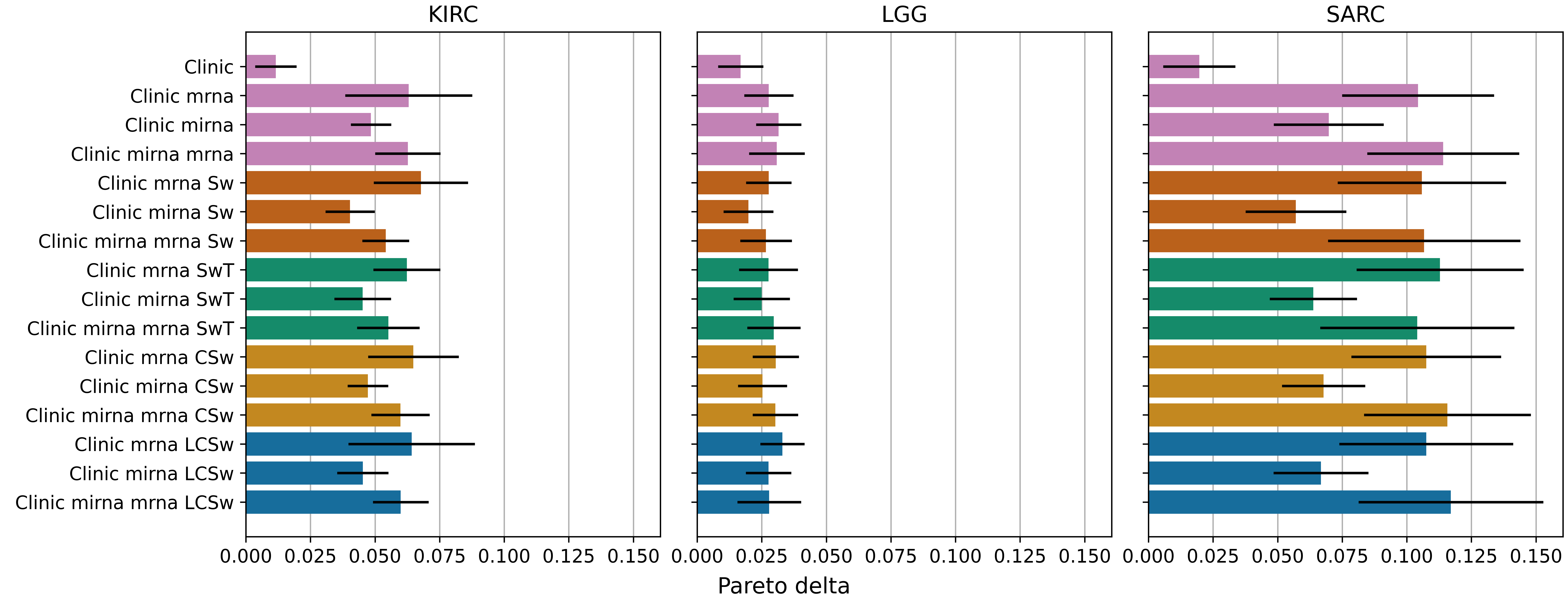}
    }

    \caption{
CHV and $P_{\Delta}$ across TCGA cohorts.
(a) CHV for each multi-omic feature selection strategy evaluated
in this study across TCGA-KIRC, TCGA-LGG, and TCGA-SARC. CHV is a MO measure of overall performance that
can be applied to cross-validation situations.
The objectives here are the predictive performance and the biomarker set size (number of selected features).
Higher values indicate better overall utility of solution sets.
(b) $P_{\Delta}$ measures the discrepancy between training and testing performance of the solutions
across the whole approximated Pareto front, when objectives
are based on ML performance metrics. $P_{\Delta}$ quantifies overestimation,
with lower values indicating smaller train–test gaps and more robust performance.
Error bars represent variability across cross-validation folds.
}
    \label{fig:chv_and_pd}
\end{figure*}

To evaluate the behavior of each multi-omic feature selection strategy,
we assess both overall performance and generalization stability (\figRef{chv_and_pd}).
Overall performance with respect to the multiple objectives is quantified using CHV.
CHV is a generalization of the hypervolume metric. The hypervolume reflects the extent of the objective space spanned
by the identified Pareto-optimal solutions. CHV generalizes the hypervolume to be applicable in a cross-validation context,
where there is a difference between expected performance after optimization
and performance measured on new data \citep{Cattelani2024IEEE}.
In this study, the objective space is defined by predictive performance and biomarker set size;
thus, CHV captures the diversity and breadth of attainable trade-offs between accuracy
and model complexity rather than peak predictive performance alone.
Overestimation is assessed using $P_{\Delta}$,
which measures the discrepancy between training and testing performance of the solutions
across the whole approximated Pareto front \citep{Cattelani2024BiB}.
As shown in \figRef{chv}, CHV varies more substantially across cohorts than across optimization strategies within each cohort.
TCGA-LGG consistently achieves the highest CHV values, indicating broader coverage of performance-complexity trade-offs.
Within LGG, concatenation-based and sweeping-based strategies yield comparable hypervolume values,
with only modest differences between them. In TCGA-KIRC, CHV values are highly similar across strategies,
and variability ranges overlap extensively, suggesting that the attainable trade-off space is captured to a comparable extent by all approaches.
In TCGA-SARC, CHV values are uniformly lower across all methods, and strategies cluster tightly with minimal separation.
This pattern reflects a narrower observed range of attainable performance-complexity
trade-offs rather than a clear distinction between optimization designs.
$P_{\Delta}$ (\figRef{pd}) further characterizes overestimation behavior.
In TCGA-LGG and TCGA-KIRC, train-test discrepancies are moderate and broadly comparable across strategies,
with no consistent increase associated with sweeping relative to direct concatenation.
In TCGA-SARC, larger $P_{\Delta}$ values are observed across all methods,
indicating stronger train-test divergence in this cohort.
These differences appear primarily cohort-driven rather than strategy-driven.
Clinical-only models consistently exhibit very small $P_{\Delta}$ values across cohorts,
reflecting lower train-test discrepancy.
In contrast, the inclusion of omic features increases model expressivity and enables the identification
of more complex composite biomarker panels, but this added flexibility is accompanied by
greater variance and larger generalization gaps.
Importantly, sweeping-based strategies do not show a systematic increase in $P_{\Delta}$ relative to direct concatenation,
indicating comparable performance stability behavior under the evaluated settings.

\subsection{Balancing predictive accuracy and biomarker set size}

\myBigFig[1.0]{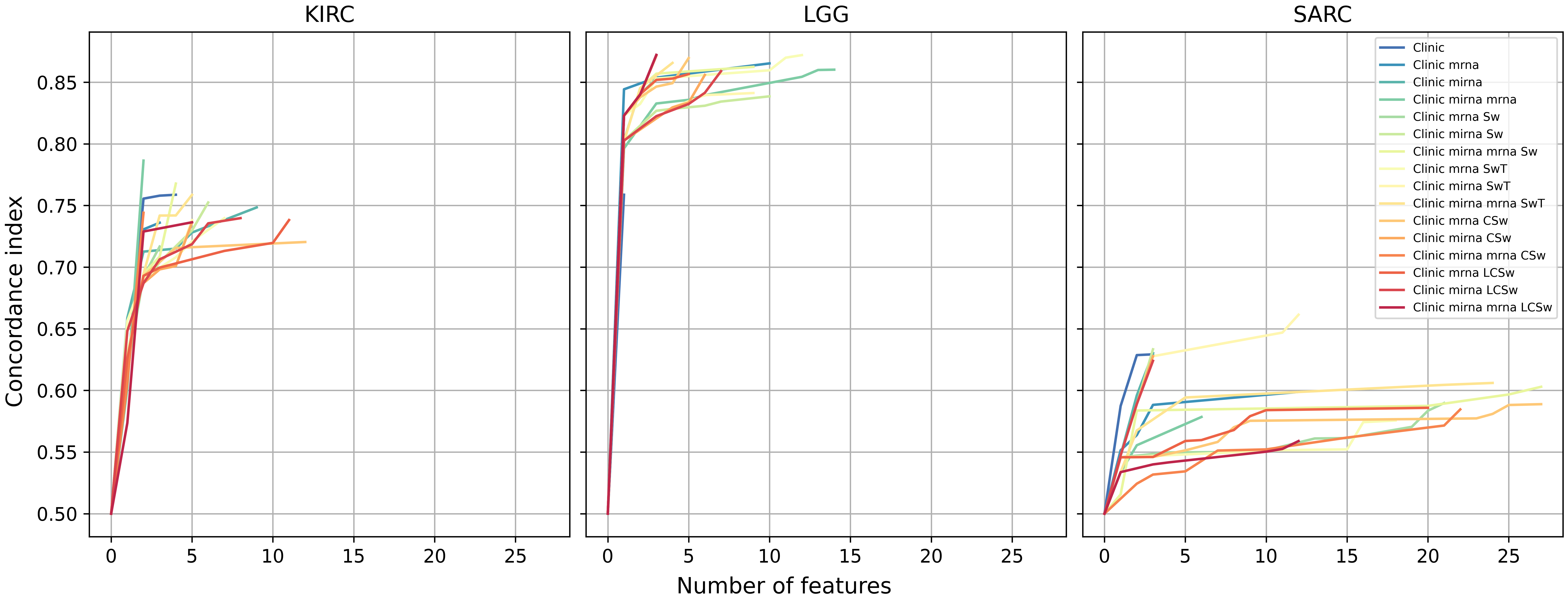}{
Accuracy-complexity trade-offs across TCGA cohorts. C-index as a function of the number of selected features
for each multi-omic feature selection strategy in TCGA-KIRC, TCGA-LGG, and TCGA-SARC.
Each curve represents the best average performance achieved across cross-validation folds at a given feature count,
illustrating the empirical trade-off between predictive accuracy and biomarker set size.
Strategies include clinical-only models, direct concatenation of omic layers, and sweeping-based MV optimization variants.
The figure highlights how different optimization designs navigate the accuracy–sparsity spectrum under distinct survival data regimes.
}

In this section, we assess whether multi-omic optimization strategies can identify compact biomarker
panels without sacrificing predictive performance. Specifically, we examine whether coordinated MV optimization (sweeping)
enables more efficient use of molecular information compared to direct concatenation,
with implications for cost-aware biomarker development.
To address this, we directly analyze the empirical trade-offs between C-index and the number of selected features achieved by each strategy.
\figRef{trade_plots} reports these accuracy-complexity relationships across cohorts,
showing C-index as a function of biomarker set size for each multi-omic feature selection approach.
The trade-off plots are computed as defined in \cite{Cattelani2024IEEE}, with C-index used in place of balanced accuracy.
In TCGA-KIRC, direct concatenation achieves concordance values that are comparable to,
and occasionally slightly higher than, those obtained by sweeping variants at similar model sizes.
The curves overlap substantially, and no strategy consistently dominates,
indicating that joint optimization over concatenated features captures most of the available survival signal in this cohort.
In TCGA-LGG, sweeping-based strategies more frequently attain similar or higher concordance with fewer selected features.
In several regions, comparable accuracy is achieved at reduced model complexity.
The separation between strategies is more apparent than in KIRC,
suggesting that sweeping-based MV optimization can yield more efficient trade-offs when the survival signal is strong.
In TCGA-SARC, concordance values are lower overall and increase gradually with model size.
Differences between strategies are modest and variable,
and no consistent advantage is observed.
Overall, the benefit of sweeping appears to depend on the survival data regime.
When sufficient event information is available, as in TCGA-LGG, sweeping can identify more compact and competitive
multi-omic biomarker panels. In more constrained settings, such as TCGA-KIRC and especially TCGA-SARC,
its performance is largely comparable to direct concatenation.

\subsection{Incremental prognostic value of multi-omic biomarker panels under nested cross-validation}

\begin{figure*}[ht!]
    \centering
    
    \subfloat[KIRC baseline best comparison.\label{fig:kirc}]{
        \includegraphics[width=0.47\textwidth]{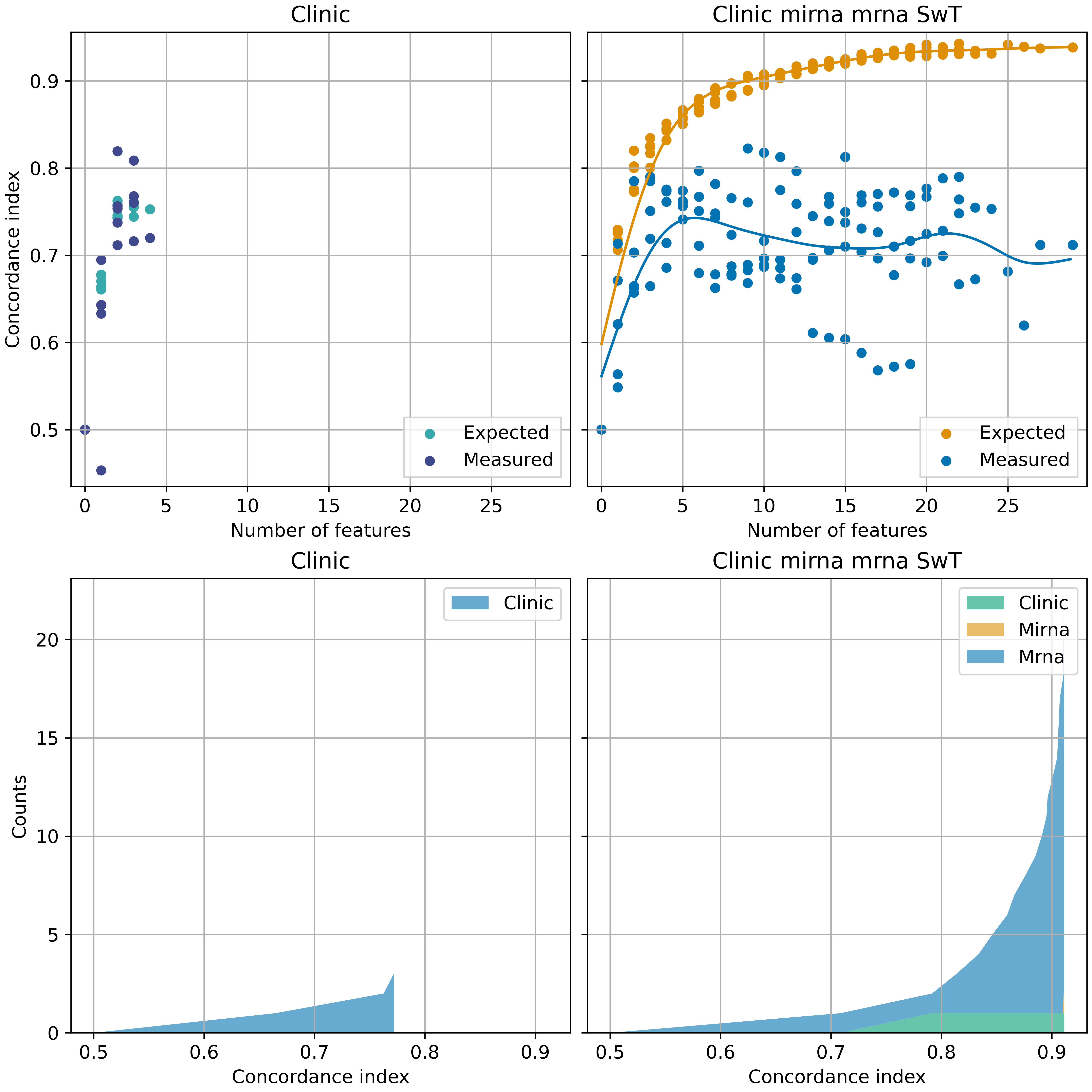}
    }\hfill
    \subfloat[LGG baseline best comparison.\label{fig:lgg}]{
        \includegraphics[width=0.47\textwidth]{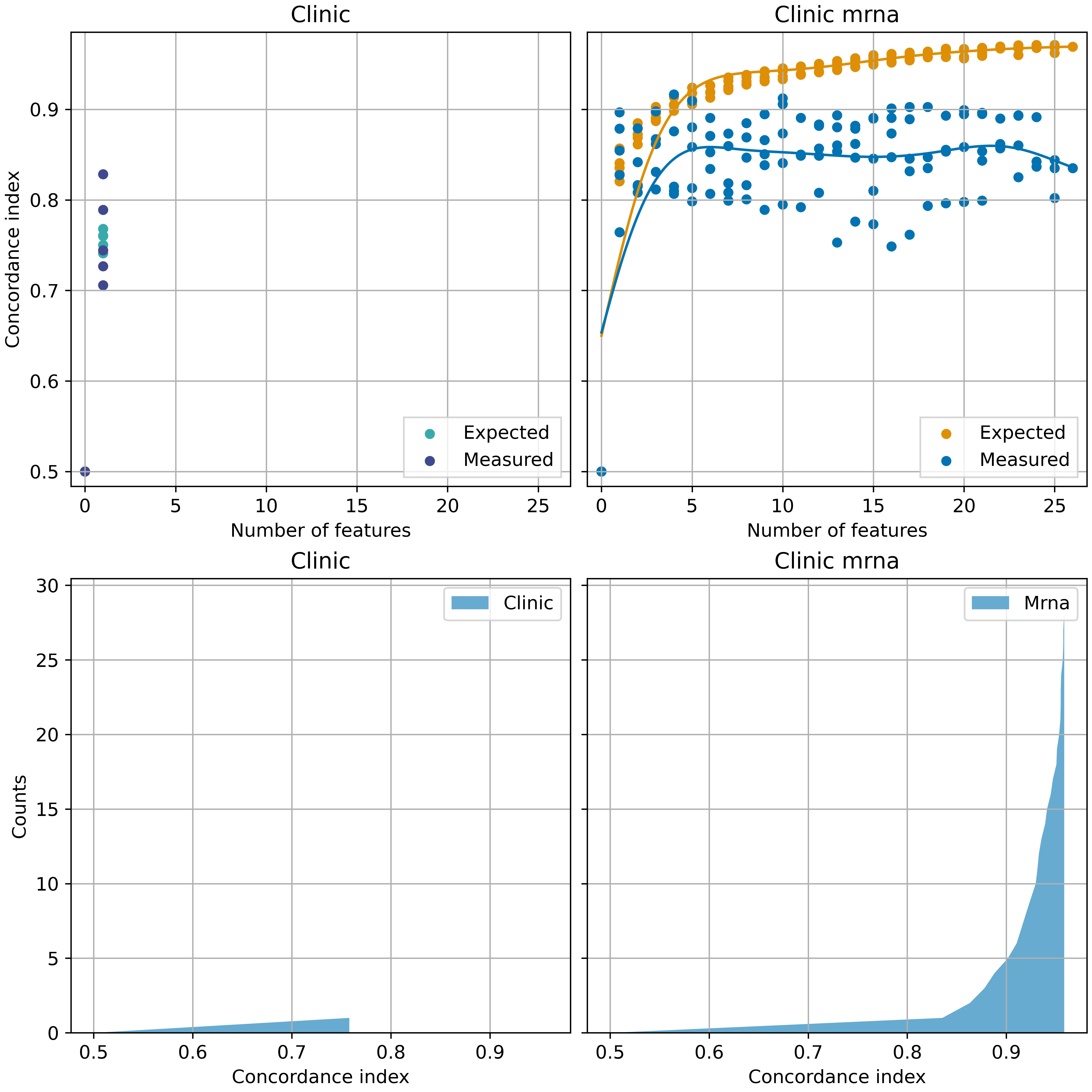}
    }

    \caption{
    	Expected versus measured C-index comparison of clinical-only and multi-omic feature selection strategies
    	in TCGA-KIRC and TCGA-LGG. Top panels show C-index as a function of the number of selected features.
    	“Expected” points represent internal cross-validation performance, while “measured” points denote testing
    	performance on the left-out sets. The results from all the 5 folds are plotted together.
    	Left column corresponds to clinical-only optimization, and right column
    	to the multi-omic sweeping strategy integrating clinical with expression-based features.
    	Bottom panels display the distribution of selected feature counts required to achieve increasing concordance levels,
    	illustrating model complexity and composition.
    	They show the counts and the expected C-index resulting from optimizing on the whole dataset.
    	The comparison highlights cohort-dependent differences in incremental molecular value and generalization stability.
}
    \label{fig:combined}
\end{figure*}

To evaluate the incremental value of molecular features beyond clinical variables,
we compared nested internal (“expected”) and external (“measured”) performance profiles.
\figRef{combined} contrasts the clinical-only baseline obtained through optimization of clinical variables
alone with a multi-omic strategy integrating clinical, and omics-driven features using the SwT configuration of the Sweeping* algorithm
in TCGA-KIRC (\figRef{kirc}) and the concatenated in TCGA-LGG (\figRef{lgg}).
These multi-omic strategies have been selected by highest CHV (\figRef{chv}).
Analogous plots for TCGA-SARC are in Supplementary Fig. S4.
In TCGA-KIRC, the clinical-only baseline reaches a stable concordance plateau with a small number of variables.
Although the multi-omic sweeping approach achieves substantially higher internal concordance as additional molecular
features are introduced, the corresponding external performance does not exceed the clinical baseline.
This indicates that the apparent internal gains primarily reflect increased variance and selection-induced optimism
rather than reproducible molecular signal in this cohort. Still, the overall performance of the multi-omic approach is
measured higher because of overestimated choices emerging by the clinic only optimization, in particular one clinic predictor
achieves a C-index of less than 0.5 on new data.
In contrast, in TCGA-LGG the clinic+mRNA strategy improves test concordance beyond the clinical baseline
across a range of feature counts. While an expected-measured gap remains, the persistence of improved measured
performance supports genuine incremental prognostic value from the molecular layer in this setting.
Feature-count distributions further show that higher expected performance levels are associated with increasingly complex
and mixed multi-omic panels. A key property of the genetic framework is that multi-omic integration is not enforced
but arises from the data. Because both the MV GAs used, the concatenated and the resampled (Supplementary Section 1.1.3),
explore also solutions that are totally missing some of the views,
the algorithm can naturally exclude an entire omic view if it does not improve the accuracy-complexity trade-off.
This implicit view-selection mechanism ensures that multi-omic biomarker panels are data-driven rather than structurally imposed,
allowing a more unbiased assessment of cross-omic contribution across cohorts.

The most frequently selected features for each Sweeping* configuration, after optimizing on the whole dataset,
are reported for TCGA-KIRC (Supplementary Fig. S5),
TCGA-LGG (Supplementary Fig. S6), and TCGA-SARC (Supplementary Fig. S7).

\section{Conclusions}

In this work, we evaluated multi-omic feature selection strategies for cancer survival prediction
using a GA framework coupled with MO optimization.
By jointly optimizing predictive performance and biomarker set size,
the proposed sweeping design enables coordinated yet data-driven integration of clinical and molecular layers
without enforcing their inclusion. Our results demonstrate that the added value of multi-omic biomarker panels
is cohort-dependent: when sufficient survival signal is present, coordinated MV optimization implemented
with alernative sweeping-based strategies can identify compact and competitive panels,
whereas in more constrained settings gains seem to be limited by data.
Importantly, the framework naturally performs both feature and view selection, providing an unbiased assessment
of cross-omic contribution. Together, these findings highlight the potential of combining evolutionary search with
MO learning for structured biomarker discovery, while underscoring the need for further research to
reduce overestimation in high-dimensional settings, and validate multi-omic panels in larger and more diverse cohorts.

\section*{Data and source code availability}
The original clinical and omic data used in this study is from public repositories.
The preprocessed data, source code, and detailed numerical results are available in a
public server (\url{github.com/UEFBiomedicalInformaticsLab/BIODAI/tree/main/Sweeping}).

\section*{Competing interests}
No competing interest is declared.

\section*{Author contributions statement}

L.C.: Conceptualization, Methodology, Software, Formal analysis, Investigation, Data curation, Validation,
Visualization, Writing – original draft, Writing – review \& editing.
V.F.: Conceptualization, Methodology, Supervision, Writing – review \& editing, Funding acquisition.

\section*{Acknowledgments}

This work was supported by the Academy of Finland (grant agreements 358037 and 373820).
The computational analyses were performed on servers provided
by the UEF Bioinformatics Center, University of Eastern Finland.

\bibliography{reference}

\end{document}


\title{\paperTitle{}, supplementary material}

\author[1]{Luca Cattelani\thanks{Corresponding Author, e-mail: luca.cattelani@uef.fi, https://orcid.org/0000-0003-4852-2310}}
\author[1]{Vittorio Fortino\thanks{Corresponding Author, e-mail: vittorio.fortino@uef.fi, https://orcid.org/0000-0001-8693-5285}}
\affil[1]{Institute of Biomedicine, School of Medicine, University of Eastern Finland, Kuopio, 70211, Finland.}
\date{}

\renewcommand\Affilfont{\small}

\maketitle

\begin{multicols}{2}

\section{Supplementary materials and methods}

\subsection{Multi-view algorithm}\label{sec:algorithm}

\mySupFig[0.6]{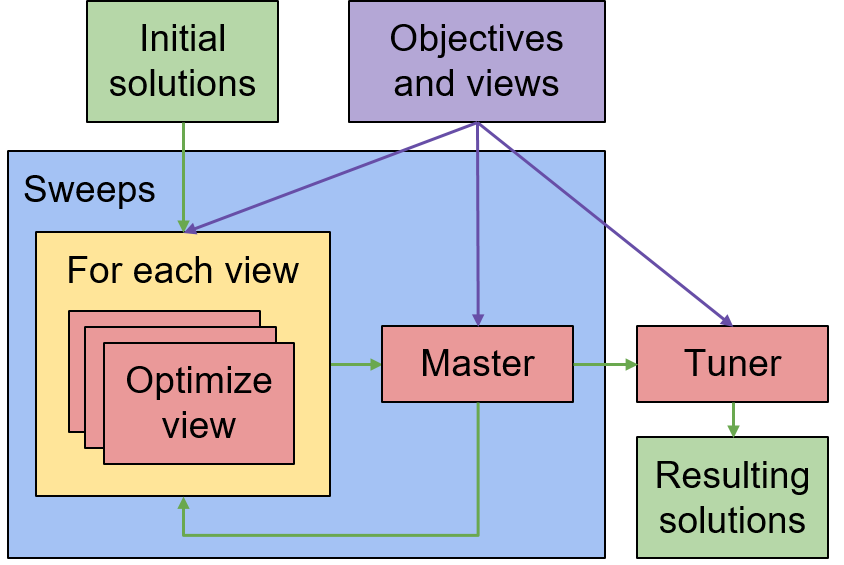}{Schematic representation of the Sweeping* algorithm.
The algorithm begins with a set of initial solutions, which may be
provided by the user or empty (in which case the algorithm starts from scratch).
These solutions form the starting populations for each view.
The user specifies objectives, i.e., evaluation metrics
(functions mapping a solution to a numerical value, possibly data‑dependent),
and views, where each view corresponds to a feature subset representing a data modality for the same sample set.
These define the problem the algorithm will optimize.
A sweep is the main unit of iteration. Each sweep consists of:
a SV step, “for each view → optimize view”
(for every view, an SV optimizer with warm start is applied),
and a MV step, “master” optimizer
(after updating all views individually, master MV optimizer jointly optimizes across all views).
The SV → MV cycle is repeated for a user‑defined number of sweeps.
After all sweeps are completed,
a tuner MV optimizer further refines the final MV population.
The tuner may also be the identity function, making this phase optional.
This final MV optimization step produces the resulting solutions.
}

In this subsection, the Sweeping* algorithm gets formally defined. We define also an extension of NSGA*,
NSGA* with Warm Start (NSGA*WS) that makes it usable as inner MV optimizer for Sweeping* by allowing it
to start from a previously defined set of solutions. This allows the chaining of multiple optimization phases as needed
by Sweeping*. A schematic representation of the Sweeping* algorithm is available in \supFigRef{sweeping_steps},
while an introductory description of the algorithm is in Section 2.

Given a fixed set of samples observed through multiple views (modalities) and a set of user‑defined objectives,
Sweeping* searches for MV solutions that achieve good trade‑offs across those objectives.
The algorithm alternates between (i) optimizing within each view separately and (ii) optimizing across all views jointly.

Sweeping* starts from a set of initial solutions (this set can be empty to start from scratch), and requires
a user provided set of objectives and of views. An objective is a function from a solution and a set of samples to
a numeric value, i.e. it is a definition of an evaluation metric that can potentially depend on a set of data-points.
A view has a set of features and assigns at every sample a value for each of its features. All the views must
have the same set of samples.
A population of MV solutions is represented by a sequence of SV solutions for each view. After a MV optimization the SV solutions at the same
position in the sequences are considered as part of the same MV solution, i.e. the MV solution has a feature set that is the union of the features of the
corresponding SV solutions.
After the SV step, instead, the order of the SV solutions in their sequence has no
meaning.
While optimizing, Sweeping* has two phases: sweeping and tuning.
During the sweeping phase, the optimizer performs a user-defined
number of sweeps each composed of a SV and a MV steps.
In the SV step, for each view, a SV MO optimizer
optimizes the population of solutions starting from its previous composition.
In the MV step, a MV optimizer considers all the views at the same time, and builds a population of MV solutions starting from the SV solutions
of the previous step. This new MV population is optimized, then either passed back to start a new SV step or used as input for the tuning phase after
the last sweep. 
The tuning phase applies an additional MV optimizer to the output of the sweeping phase.
The output of the tuner algorithm is used as the result of the whole
Sweeping* algorithm. Every optimizer can implement the identity function (return the initial population) so every part of Sweeping* is
in fact optional. Which SV algorithms
to use for each view and each sweep, which master algorithm to use for each sweep, and which tuning algorithm to run at the end,
are user-defined hyperparameters.

\subsubsection{\textsc{Sweeping*}}\label{sec:sweeping}

\supAlgorithmStart{\textsc{Sweeping*}}{sweeping_star}
	\State \textbf{class} \textsc{Sweeping*} \textbf{inherits} \textsc{MVMOOptimizer}
		\State \tab{} \textbf{method} \textsc{new}($sweepingStrategy$)
    			\State \tab{}\tab{} $self.sweepingStrategy \gets sweepingStrategy$
		\State \tab{} \textbf{end method}
		\State \tab{} \textbf{method} \textsc{optimizeMV}($objectives, views, pops$)
			\State \tab{}\tab{}\textbf{for} $s$ in \Call{$sweepingStrategy.sweeps$}{} \textbf{do}
				\State \tab{}\tab{}\tab{}\textbf{for} $i$ in $0:\left| views \right|$ \textbf{do}
					\State \tab{}\tab{}\tab{}\tab{} $pops[i] \gets$ \Call{$s.optimizers[i]$.optimizeWS}{\raggedright{}\linebreak{}
						\tab{}\tab{}\tab{}\tab{}\tab{}\tab{}$objectives$, $views[i]$, $pops[i]$}
				\State \tab{}\tab{}\tab{}\textbf{end for}
				\State \tab{}\tab{}\tab{} $pops \gets$ \Call{$s.master$.optimizeMV}{\raggedright{}\linebreak{}
					\tab{}\tab{}\tab{}\tab{}\tab{}$objectives$, $views$, $pops$}
	    		\State \tab{}\tab{}\textbf{end for}
	    		\State \tab{}\tab{} $tuner \gets$ \Call{$sweepingStrategy.tuner$}{}
	    		\State \tab{}\tab{} \Return \Call{$tuner$.optimizeMV}{$objectives$, $views$, $pops$}
		\State \tab{} \textbf{end method}
	\State \textbf{end class}
\supAlgorithmEnd{}

The \textsc{Sweeping*} is an \textsc{MVMOOptimizer} (\supSecRef{mvmoo}) that is customized at creation with a $sweepingStrategy$ object.
While optimizing, \textsc{Sweeping*} has two phases: sweeping and tuning. During the sweeping phase, the optimizer performs a
number of sweeps as defined in the $sweepingStrategy$. For each sweep, there are a SV and a MV steps.
In the SV step, for each view,
a SV MO optimizer with warm start (\textsc{MOOptimizerWS}, \supSecRef{mooptwarmstart})
optimizes the population starting from its previous composition. In the MV step, a MV optimizer is used
to try to improve all the populations considered at the same time using all of views. After all the sweeps have been performed,
a tuning MV optimizer may be executed, using the current populations as its initial ones. Which SV algorithms
to use for each view and each sweep, which master algorithm to use for each sweep, and which tuning algorithm to run at the end,
are hyperparameters specified through the $sweepingStrategy$ object. A pseudocode definition of the \textsc{Sweeping*} algorithm is
in \supCodeRef{sweeping_star}.
\textsc{Sweeping*} has the following properties.
\begin{itemize}
    \item It accepts initial populations in input, but can also start from scratch if empty initial populations are passed.
    \item It is a generalization of any \textsc{MVMOOptimizer}, since the behaviour of any \textsc{MVMOOptimizer} can be
    reproduced by setting it as tuner, and setting 0 sweeps in the $sweepingStrategy$ object.
    \item The tuner can implement the identity function returning the same populations it receives in input.
    Consequently, the tuning phase is optional.
    \item Master optimizers and tuner can potentially be other \textsc{Sweeping*} optimizers, allowing for complex composite
    strategies.
\end{itemize}

\mySupFig{class_diagram}{
UML class diagram that visually summarizes the main software components used to implement the \textsc{Sweeping*}
algorithm and their dependencies.
The classes fall into several conceptual groups, each playing a specific role in the framework.
The core optimizer classes that are the fundamental building blocks for optimization:
\textsc{MOOptimizer} (the base class for all SV MO optimizers,
it extends \textsc{MOOptimizerWS} to support MV optimization with warm start),
\textsc{Sweeping*} (a composite optimizer implementing \textsc{MVMOOptimizer}),
\textsc{NSGA*WS} (a warm‑start version of the NSGA* algorithm,
can be used both as SV and MV optimizer inside the \textsc{Sweeping*} framework).
Warm‑start infrastructure
(warm‑starting allows optimizers to begin from existing populations, enabling pipelines):
\textsc{MOOptimizerWS} (extends \textsc{MOOptimizer}
to add an \textsc{optimizeWS} method that accepts an initial population),
\textsc{MOOptimizerWSFactory} and \textsc{NSGA*Factory}
(these factory classes create warm‑start optimizers on demand, using the current populations).
Genetic algorithm components,
these classes determine how populations evolve during optimization:
\textsc{GAStrategy} (an abstract class encapsulating GA‑specific behaviors),
two concrete implementations are shown (
\textsc{GASBitList}, bitlist‑based feature encoding,
\textsc{GASSelector}, MV individuals composed of SV individuals, one per view),
\textsc{GASFactory}, \textsc{GASFactoryBitList}, \textsc{GASFactorySelector} 
(factories for creating \textsc{GAStrategy} instances tailored to the current population handling strategy).
\textsc{MVMOOptimizer} implementations with specific
MV composition strategies, these define how separate views are combined to form MV individuals:
\textsc{Concatenated} (concatenates SV feature sets into a single large feature set,
possibly with probabilistic feature retention in the lean configuration),
\textsc{Resampled} (uses a GA where each MV individual is formed by one SV individual per view,
handles MV structure through resampling rather than concatenation).
Together, these classes form a modular and extensible architecture for MV MO optimization.
}

The Sweeping Algorithm is based on wrapping and composing of simpler SV algorithms.
To develop this framework, the following elements were needed.
\begin{itemize}
    \item MO algorithms supporting ``warm start'': the possibility to start from a previous population. In this manner pipelines of optimizers can be formed.
    \item Parts of the pipelines use SV solution parts and a resampling strategy (see Section 1.1.3). To obtain this behavior,
    GA parts must be customized on the initial population, which in turn requires to create the optimizers on the fly using factory objects.
    \item To reduce the clutter, the parts of the GA algorithm that can change are encapsulated in strategy objects (see \supSecRef{gastrategy}).
\end{itemize}
\supFigRef{class_diagram} shows the main classes that are used to compose the MV MO algorithm that is featured in our tests and their dependencies.
The classes will be described in the following.

\subsubsection{Multi-objective optimizer with warm start}\label{sec:mooptwarmstart}

\supAlgorithmStart{\textsc{MOOptimizerWS} abstract class definition}{mooptws}
    \State \textbf{class} \textsc{MOOptimizerWS} \textbf{inherits} \textsc{MOOptimizer}
		\State \tab{} \textbf{method} \textsc{optimizeWS}($objectives, trainingData, pop$)
    		\State \tab{} \textbf{method} \textsc{optimize}($objectives, trainingData$)
    			\State \tab{}\tab{}\Return \Call{optimizeWS}{$objectives, trainingData, \emptyset{}$}
    		\State \tab{} \textbf{end method}
    \State \textbf{end class}
    \State \textbf{class} \textsc{MOOptimizerWSFactory}
		\State \tab{} \textbf{method} \textsc{createWS}($pops$)
    \State \textbf{end class}
\supAlgorithmEnd{}

In order to compose multiple MO optimizers in a pipeline, we need optimizers able to start from an initial
population that is the result of a previous optimization.
The class \textsc{MOOptimizerWS} extends \textsc{MOOptimizer} allowing to warm-start the optimization process with
an initial population of solutions. Calling \textsc{optimizeWS} with an empty set of solutions is equivalent to
use the optimizer without a warm start. For reasons that will became obvious in the following, in order to use
SV optimizers as building blocks of MV optimizers, there is the need to include the design
scenarios in which the inner workings of a SV optimizer, like for example the genetic operators of a GA,
are affected by an initial population or even multiple initial populations. To take this into account an
\textsc{MOOptimizerWSFactory} class is provided. Its instantiations can create \textsc{MOOptimizerWS} objects on the fly taking
into account initial populations. The case in which there is only one initial population is simply handled by passing a
collection containing a single population to the method \textsc{createWS}.
The pseudocode definitions of \textsc{MOOptimizerWS} and \textsc{MOOptimizerWSFactory} are in \supCodeRef{mooptws}.

\subsubsection{Multi-view multi-objective optimizers}\label{sec:mvmoo}

\supAlgorithmStart{\textsc{MVMOOptimizer} abstract class definition}{mvmoopt}
    \State \textbf{class} \textsc{MVMOOptimizer} \textbf{inherits} \textsc{MOOptimizerWS}
		\State \tab{} \textbf{method} \textsc{optimizeMV}($objectives, views, pops$)
    		\State \tab{} \textbf{method} \textsc{optimizeWS}($objectives, trainingData, pop$)
    			\State \tab{}\tab{}\Return \Call{optimizeMV}{\raggedright{}\linebreak{}
    			\tab{}\tab{}\tab{}\tab{} $objectives, [trainingData], [pop]$}$[0]$
    		\State \tab{} \textbf{end method}
    \State \textbf{end class}
    \State \textbf{class} \textsc{Concatenated} \textbf{inherits} \textsc{MVMOOptimizer}
    		\State \tab{} \textbf{method} \textsc{new}($optimizerFactory, lean$)
    			\State \tab{}\tab{} $fact \gets optimizerFactory$
    			\State \tab{}\tab{} $self.lean \gets lean$
		\State \tab{} \textbf{end method}
		\State \tab{} \textbf{method} \textsc{optimizeMV}($objectives, views, pops$)
			\State \tab{}\tab{} $cViews \gets$ \Call{concat}{$views$}
			\State \tab{}\tab{} $opt \gets$  \Call{$fact$.create}{$pops$}
			\State \tab{}\tab{} $pop \gets$ \Call{concatenatePops}{$lean, pops$}
			\State \tab{}\tab{} $pop \gets$ \Call{$opt$.optimizeWS}{$objectives, cViews, pop$}
			\State \tab{}\tab{}\Return \Call{viewPopsFromConcatenated}{$pop$}
		\State \tab{} \textbf{end method}
    \State \textbf{end class}
    \State \textbf{class} \textsc{Resampled} \textbf{inherits} \textsc{MVMOOptimizer}
    		\State \tab{} \textbf{method} \textsc{new}($optimizerFactory$)
    			\State \tab{}\tab{} $fact \gets optimizerFactory$
		\State \tab{} \textbf{end method}
		\State \tab{} \textbf{method} \textsc{optimizeMV}($objectives, views, pops$)
			\State \tab{}\tab{} $cViews \gets$ \Call{concat}{$views$}
			\State \tab{}\tab{} $opt \gets$  \Call{$fact$.create}{$pops$}
			\State \tab{}\tab{} $pop \gets$ \Call{$opt$.optimize}{$objectives, cViews$}
			\State \tab{}\tab{}\Return \Call{viewPopsFromSampling}{$pop$}
		\State \tab{} \textbf{end method}
    \State \textbf{end class}
\supAlgorithmEnd{}

\textsc{MVMOOptimizer} objects are \textsc{MOOptimizerWS} objects that can also be applied to MV situations.
Optimization with a single view is a special case of optimization with multiple views, and can be adapted for a
MV optimizer simply by calling the method for MV optimization, \textsc{optimizeMV}, with
a collection containing a single view and another collection containing a single initial population. For this work, we needed two implementations of
\textsc{MVMOOptimizer}: \textsc{Concatenated} and \textsc{Resampled}. These are two strategies to handle MV scenarios by using
SV algorithms internally. In \textsc{Concatenated}, the views are concatenated
(and also the initial populations, if the warm-start feature is used),
forming a single view with as features the union of the features of all the views,
then a SV algorithm with warm start is applied to the dataset resulting from the concatenation.
Afterwards, a result population for each view is obtained by splitting each concatenated solution
so that for each view there is an individual with only the features of the concatenated individual that belong to that view.
The initial population is created by permuting randomly the populations of each view, then concatenating the individuals at the same position.
In the $lean$ variant of \textsc{Concatenated}, after creating an individual by concatenation of an individual from each view, it is filtered by keeping each feature
with a probability equal to the proportion of its view in the set of features of that individual.
The \textsc{Resampled} \textsc{MVMOOptimizer} also concatenates the views, but is radically different in the way it handles the individuals.
An \textsc{MOOptimizerWSFactory} creates an optimizer according to an initial population for each view, so that the internal operations of the
optimizer, like the creation of the initial solutions, mutation and crossover operators, can depend on the initial populations.
Each individual of the \textsc{Resampled} optimizer is formed by a union of a SV individual for each view. At the end of the execution the result populations are
formed by simply separating each MV individual in its components and putting them in the result collection of the corresponding view.
The pseudocode definitions of \textsc{MVMOOptimizer}, \textsc{Concatenated}, and \textsc{Resampled} are in \supCodeRef{mvmoopt}.

\subsubsection{Genetic algorithm strategy}\label{sec:gastrategy}

\supAlgorithmStart{\textsc{GAStrategy} abstract class definition}{gastrategy}
    \State \textbf{class} \textsc{GAStrategy}
    		\State \tab{} \textbf{method} \textsc{featureImportance}($objectives, trainData$)
    		\State \tab{} \textbf{method} \textsc{createNewIndividuals}($featImp, num$)
		\State \tab{} \textbf{method} \textsc{evaluate}($pop, objectives, trainData$)
    		\State \tab{} \textbf{method} \textsc{sort}($pop$)
    		\State \tab{} \textbf{method} \textsc{crossover}($pop$)
    		\State \tab{} \textbf{method} \textsc{mutation}($pop$)
    		\State \tab{} \textbf{method} \textsc{handleClones}($pop, featImp$)
    \State \textbf{end class}
    \State \textbf{class} \textsc{GASBitlist} \textbf{inherits} \textsc{GAStrategy}
    	\State \tab{} \dots
    \State \textbf{end class}
    \State \textbf{class} \textsc{GASSelector} \textbf{inherits} \textsc{GAStrategy}
    	\State \tab{} \dots
    \State \textbf{end class}
\supAlgorithmEnd{}

The \textsc{GAStrategy} abstract class is used to encapsulate some aspects of a GA (e.g. what mutation operator to use),
so that they can be polymorphically modified without having to redefine the main structure of the GA.
A \textsc{GAStrategy} defines how to compute a feature importance, how to create new individuals, how to evaluate them,
how to sort them in an order of preference, what kind of crossover and mutation operators to use, and how to handle clones
(individuals with the same set of features).
For this work, we used two kinds of \textsc{GAStrategy}: \textsc{GASBitlist} and \textsc{GASSelector}.
They differ mainly in the way they represent the individuals. \textsc{GASBitlist} conceptually represents individuals as lists of bits where
each position represents a feature and a ``1'' in a position indicates that the feature at that position is active in the individual
(in practice we use sparse lists to save memory). For the experiments described in this work,
the operations for \textsc{GASBitlist} are implemented as in the NSGA3-CHS MO optimizer instantiation previously described in \cite{Cattelani2024BiB}.
\textsc{GASSelector} manages MV individuals that are composed by SV individuals, one for each view. When a \textsc{GASSelector}
object is instantiated it memorizes a vocabulary of individuals for each view.
When a new MV individual is created, a random SV individual is extracted from each vocabulary.
The mutation operator can switch a single view individual with
another one (with a probability of $1/|views|$ in our experiments).
The crossover operator forms a new MV individual by extracting the SV individuals from the MV parents, with equal probability.
Feature importance is not used (calling the method just returns the uniform distribution).
The pseudocode definition of \textsc{GAStrategy}, and the signatures of \textsc{GASBitlist} and \textsc{GASSelector}, are in \supCodeRef{gastrategy}.

\supAlgorithmStart{\textsc{GASFactory} abstract class definition}{gastrategyFact}
    \State \textbf{class} \textsc{GASFactory}
    		\State \tab{} \textbf{method} \textsc{create}($pops$)
    \State \textbf{end class}
    \State \textbf{class} \textsc{GASFactoryBitlist} \textbf{inherits} \textsc{GASFactory}
    	\State \tab{} \dots
    \State \textbf{end class}
    \State \textbf{class} \textsc{GASFactorySelector} \textbf{inherits} \textsc{GASFactory}
    	\State \tab{} \dots
    \State \textbf{end class}
\supAlgorithmEnd{}

\textsc{GASFactory} objects are used to create \textsc{GAStrategy} objects on the fly, using current information about the population of solutions for each view.
In the context of this work, we used two types of \textsc{GASFactory}: \textsc{GASFactoryBitlist} to create \textsc{GASBitlist} objects,
and \textsc{GASFactorySelector} to create \textsc{GASSelector} objects. The constructor methods of the \textsc{GASFactory} concrete classes receive
and memorize all the necessary parameters, like the crossover and mutation frequencies or the sorting strategy. This code is omitted here, we
refer the interested readers to the source code public repository.
The pseudocode definition of \textsc{GASFactory},
and the signatures of \textsc{GASFactoryBitlist} and \textsc{GASFactorySelector}, are in \supCodeRef{gastrategyFact}.

\subsubsection{\textsc{NSGA*} with warm start}

\supAlgorithmStart{\textsc{NSGA*WS} class definition}{nsgaStarWS}
    \State \textbf{class} \textsc{NSGA*WS} \textbf{inherits} \textsc{MOOptimizerWS}
    		\State \tab{} \textbf{method} \textsc{new}($nGenerations, popsize, tournament,$\raggedright{}\linebreak{}
    				\tab{}\tab{}\tab{}$gaStrategy, hofFactory$)
    			\State \tab{}\tab{} $self.nGenerations \gets nGenerations$
    			\State \tab{}\tab{} $self.popsize \gets popsize$
    			\State \tab{}\tab{} $self.tournament \gets tournament$
    			\State \tab{}\tab{} $gas \gets gaStrategy$
    			\State \tab{}\tab{} $self.hofFactory \gets hofFactory$
    		\State \tab{} \textbf{end method}
    		
    		\State \tab{} \textbf{method} \textsc{optimizeWS}($objectives, trainData, pop$)
    			\State \tab{}\tab{} $hof \gets$ \Call{$hofFactory$.createNewHallOfFame}{}
    			\State \tab{}\tab{} $featImp \gets$ \Call{$gas$.featureImportance}{\raggedright{}\linebreak{}
    				\tab{}\tab{}\tab{}\tab{}$objectives, trainData$}
    			\State \tab{}\tab{} $pop[\left| pop \right|:popsize] \gets$ \raggedright{}\linebreak{}
    					\tab{}\tab{}\tab{}\tab{}\Call{$gas.$createNewIndividuals}{\raggedright{}\linebreak{}
    					\tab{}\tab{}\tab{}\tab{}\tab{}\tab{}$featImp, popsize-pop$}
    			\State \tab{}\tab{}\textbf{for} $0:nGenerations$ \textbf{do}
    				\State \tab{}\tab{}\tab{} $pop \gets$ \Call{$gas$.evaluate}{$pop, objectives, trainData$}
    				\State \tab{}\tab{}\tab{} $pop \gets$ \Call{$gas$.sort}{$pop$}
    				\State \tab{}\tab{}\tab{} $offspring \gets$ \Call{tournament}{$pop$}
    				\State \tab{}\tab{}\tab{} $offspring \gets$ \Call{$gas$.crossover}{$offspring$}
    				\State \tab{}\tab{}\tab{} $offspring \gets$ \Call{$gas$.mutation}{$offspring$}
         			\State \tab{}\tab{}\tab{} $pop \gets$ \Call{concat}{$pop, offspring$}
         			\State \tab{}\tab{}\tab{}	$pop \gets$ \Call{$gas$.handleClones}{$pop, featImp$}
         			\State \tab{}\tab{}\tab{} $pop \gets$ \Call{$gas$.evaluate}{$pop, objectives, trainData$}
         			\State \tab{}\tab{}\tab{} $pop \gets$ \Call{$gas$.sort}{$pop$}
         			\State \tab{}\tab{}\tab{} \Call{$hof.$update}{$pop$}
         			\State \tab{}\tab{}\tab{} $pop \gets pop[1..popsize]$
    			\State \tab{}\tab{}\textbf{end for}
    			\State \tab{}\tab{}\Return $hof$
    		\State \tab{} \textbf{end method}
    	\State \textbf{end class}
\supAlgorithmEnd{}

The NSGA* algorithm has been introduced in \cite{Cattelani2024BiB}. For easy of reference, we offer a rewriting of NSGA*
with the pseudocode style used in this work in \supSecRef{nsga}.
NSGA* with warm start (NSGA*WS) is a generalization of NSGA* that allows to use an initial population that is provided from outside the algorithm.
This is an essential feature to participate in pipelines of MO optimizers. \textsc{NSGA*WS} receives an memorizes at construction all the required
hyperparameters, including a \textsc{GAStrategy} object. This encapsulates the details of some operations, like e.g. the mutation operator, so that they
can be changed without redefining the whole algorithm. The \textsc{GAStrategy} also handles the representation of the individuals. Having the possibility
to run NSGA* with different representations for the individuals is crucial for the MV approach presented here. During construction, an \textsc{NSGA*WS}
also receives a factory object for hall of fames. A hall of fame is an observer of the population during the generations, that updates an internal collection
of preferred individuals and returns it at the end of the optimization.
To build the pipelines used in this work, more than one kind of hall of fame is needed.
When the optimization starts (method \textsc{optimizeWS}) the algorithm asks to the hall of fame factory object to instantiate a new hall of fame, that will
then be used throghout the generations.
It also uses the \textsc{GAStrategy} to create new individuals as needed to fill the population up to the required $popsize$.
This way it is not necessary to pass from outside an initial population of cardinality $popsize$,
and the scenario without initial population is handled as a special case.
If the passed population is bigger than $popsize$ it will be cropped at the end of the first generation when the best $popsize$ individuals are selected.
The pseudocode definition of \textsc{NSGA*WS} is in \supCodeRef{nsgaStarWS}.

\subsubsection{\textsc{NSGA*} with warm start factory}

\supAlgorithmStart{\textsc{NSGA*Factory} abstract class definition}{MOOptimizerWSFactory}
    \State \textbf{class} \textsc{NSGA*Factory} \textbf{inherits} \textsc{MOOptimizerWSFactory}
    		\State \tab{} \textbf{method} \textsc{new}($nGenerations, popsize, tournament,$\raggedright{}\linebreak{}
    				\tab{}\tab{}\tab{}$gasFactory, hofFactory$)
    			\State \tab{}\tab{} $self.nGenerations \gets nGenerations$
    			\State \tab{}\tab{} $self.popsize \gets popsize$
    			\State \tab{}\tab{} $self.tournament \gets tournament$
    			\State \tab{}\tab{} $self.gasFactory \gets gasFactory$
    			\State \tab{}\tab{} $self.hofFactory \gets hofFactory$
		\State \tab{} \textbf{end method}
		\State \tab{} \textbf{method} \textsc{create}($pops$)
			\State \tab{}\tab{} $gas \gets gasFactory.$\Call{create}{$pops$}
			\State \tab{}\tab{}\Return \Call{new $\text{NSGA*WS}$}{\raggedright{}\linebreak{}
			\tab{}\tab{}\tab{}\tab{}$nGenerations, popsize, tournament,$\raggedright{}\linebreak{}
			\tab{}\tab{}\tab{}\tab{}$gas, hofFactory$}
		\State \tab{} \textbf{end method}
    \State \textbf{end class}
\supAlgorithmEnd{}

As seen previously, \textsc{MOOptimizerWSFactory} objects are needed to construct MO optimizers taking into account the current population.
They must contain information about the desired hyperparameters.
\supCodeRef{MOOptimizerWSFactory} defines the \textsc{NSGA*Factory} factory class for \textsc{NSGA*WS} objects.

\subsection{NSGA*}\label{sec:nsga}

\supAlgorithmStart{\textsc{MOOptimizer} abstract class definition}{moopt}
    \State \textbf{class} \textsc{MOOptimizer}
    		\State \tab{} \textbf{method} \textsc{optimize}($objectives, trainingData$)
    \State \textbf{end class}
\supAlgorithmEnd{}

\supAlgorithmStart{\textsc{Nsga*} class definition}{nsgaStar}
    \State \textbf{class} \textsc{NSGA*} \textbf{inherits} \textsc{MOOptimizer}
    		\State \tab{} \textbf{method} \textsc{new}(\raggedright{}\linebreak{}
    				\tab{}\tab{}\tab{}$nGenerations, popsize, featureImportance, sort,$\linebreak{}
    				\tab{}\tab{}\tab{}$tournament, mutation, cloneRepurposing$)
    			\State \tab{}\tab{} $self.nGenerations \gets nGenerations$
    			\State \tab{}\tab{} $self.popsize \gets popsize$
    			\State \tab{}\tab{} $self.featureImportance \gets featureImportance$
    			\State \tab{}\tab{} $self.sort \gets sort$
    			\State \tab{}\tab{} $self.tournament \gets tournament$
    			\State \tab{}\tab{} $self.mutation \gets mutation$
    			\State \tab{}\tab{} $self.cloneRepurposing \gets cloneRepurposing$
    		\State \tab{} \textbf{end method}
    		
    		\State \tab{} \textbf{method} \textsc{optimize}($objectives, trainingData$)
    			\State \tab{}\tab{} $hof \gets$ \Call{createNewHallOfFame}{}
    			\State \tab{}\tab{} $featImp \gets$ \Call{$self.featureImportance$}{$trainingData$}
    			\State \tab{}\tab{} $pop \gets$ \Call{createNewRandomIndividuals}{\raggedright{}\linebreak{}
    					\tab{}\tab{}\tab{}\tab{}$featImp, self.popsize$}
    			\State \tab{}\tab{}\textbf{for} $1:self.nGenerations$ \textbf{do}
    				\State \tab{}\tab{}\tab{} $pop \gets$ \Call{evaluate}{$pop, objectives$}
    				\State \tab{}\tab{}\tab{} $pop \gets$ \Call{$self$.sort}{$pop$}
    				\State \tab{}\tab{}\tab{} $offspring \gets$ \Call{$self$.tournament}{$pop$}
    				\State \tab{}\tab{}\tab{} $offspring \gets$ \Call{crossover}{$offspring$}
    				\State \tab{}\tab{}\tab{} $offspring \gets$ \Call{$self$.mutation}{$offspring$}
         		\State \tab{}\tab{}\tab{} $pop \gets$ \Call{concat}{$pop, offspring$}
         		\State \tab{}\tab{}\tab{} \textbf{if} $self.cloneRepurposing$ \textbf{then}
         		\State \tab{}\tab{}\tab{}\tab{}	$pop \gets$ \Call{replaceClonesWithNewIndividuals}{\raggedright{}\linebreak{}
         		\tab{}\tab{}\tab{}\tab{}\tab{}\tab{} $pop, featImp$}
         		\State \tab{}\tab{}\tab{} \textbf{end if}
         		\State \tab{}\tab{}\tab{} $pop \gets$ \Call{evaluate}{$pop, objectives$}
         		\State \tab{}\tab{}\tab{} $hof \gets$ \Call{updateHallOfFame}{$hof, pop$}
         		\State \tab{}\tab{}\tab{} $pop \gets$ \Call{$self$.sort}{$pop$}
         		\State \tab{}\tab{}\tab{} $pop \gets pop[1..self.popsize]$
    			\State \tab{}\tab{}\textbf{end for}
    		\State \tab{} \textbf{end method}
    	\State \textbf{end class}
\supAlgorithmEnd{}

NSGA* has been introduced  in \cite{Cattelani2024BiB}.
In\supCodeRef{nsgaStar}, we provide a rewriting of \textsc{NSGA*}
with the pseudocode style used in this work.
Similarly, \supCodeRef{moopt} is a rewriting of the interface \textsc{MultiObjectiveOptimizer} from the same work,
now shortened to \textsc{MOOptimizer}.

\subsection{Sweeping* algorithm setups}\label{sec:setups}

\begin{table*}[tbp]
\caption{Considered configurations of the Sweeping* algorithm.\label{tab:sweepingSetups}}
\tabcolsep=0pt
\begin{tabular*}{\textwidth}{@{\extracolsep\fill}lllll@{\extracolsep\fill}}
\toprule
name & nick & master & sweeping generations & tuning generations\\
\midrule
concatenated & & & [] & 300 \\
resampled sweeping & Sw & Resampled & [50, 50, 50] & 0 \\
concatenated sweeping & CSw & Concatenated(lean=false) & [150] & 0\\
lean concatenated sweeping & LCSw & Concatenated(lean=true) & [150] & 0\\
resampled sweeping with tuning & SwT & Resampled & [50, 50] & 100\\
\bottomrule
\end{tabular*}
\end{table*}

In our benchmark, we compared five different setups of the Sweeping* algorithm.
When an SV optimizer is required, the algorithm always employs the NSGA3‑CHS inner optimizer \citep{Cattelani2024BiB},
using a population size of 500 and three inner folds for evaluating the individuals.
All other hyperparameters follow those in \cite{Cattelani2024BiB}.
The Sweeping* algorithm is introduced in \supSecRef{sweeping},
and the five configurations considered in this work are listed in \tabRef{sweepingSetups}.
Each configuration has a full name and a shorter nickname used throughout the manuscript.
The differences between the resampled, lean‑concatenated, and non‑lean‑concatenated master steps
are described in \supSecRef{mvmoo}.
Sweeping generations are expressed as sequences that indicate the number of generations performed in each sweep.
For every sweep, the specified number of generations is executed in both the SV and MV steps (see \supSecRef{sweeping}).
When a tuning phase is included, it is always performed using a concatenated NSGA3‑CHS optimizer.
Since each sweep contains both SV and MV steps, the total number of generations for every configuration is always 300.

\subsection{Case study}\label{sec:caseStudy}

\mySupBigFig[1.0]{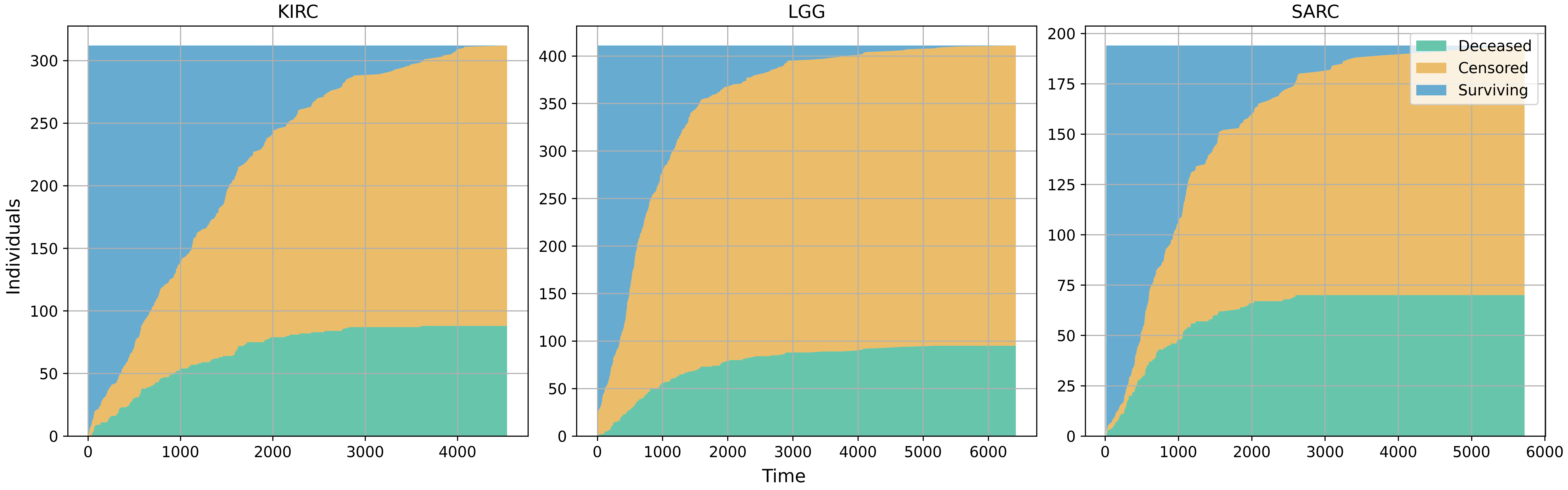}{
Event accumulation and censoring patterns across TCGA survival cohorts.
Stacked area plots show cumulative numbers of deceased, censored, and surviving individuals over follow-up time
for TCGA-KIRC, TCGA-LGG, and TCGA-SARC. The three cohorts display distinct survival data regimes,
with differences in timing and density of observed events as well as censoring proportions.
}

The considered cancer types are kidney renal clear cell carcinoma (KIRC),
brain lower grade glioma (LGG),
and sarcoma (SARC). For each cancer type we included the same three distinct data layers
from The Cancer Genome Atlas (TCGA) project \citep{hutter2018cancer} as preprocessed in \cite{Wissel2023}:
clinical information, mRNA expression, and miRNA transcription profiles.
The clinical features are age, gender, race, and tumor stage.
The number of samples for KIRC, LGG, and SARC are respectively 312, 411, and 194.
The three TCGA cohorts represent distinct survival signal regimes
that influence the feasibility of MO feature selection (\supFigRef{survival_stackplot}).
TCGA-LGG shows early and abundant event accumulation, providing a strong survival signal that supports exploration
of more complex feature combinations.
Considering the total number of events, TCGA-KIRC exhibits a slightly more gradual event accumulation,
representing an intermediate setting in which molecular features may refine clinical predictors.
In contrast, TCGA-SARC has fewer observable events,
limiting statistical power and increasing uncertainty in multi-omic modeling.
These differences provide essential context for interpreting our results,
as the ability to identify informative multi-omic feature combinations
can depend directly on the availability of survival signal (number of events) within each cohort.

We have benchmarked five setups of the sweeping algorithm
in the context of MV MO biomarker discovery for cancer survival prediction (\tabRef{sweepingSetups}).
As inner SV MO optimizer,
we used a recently proposed modification of NSGA3 \citep{Deb2014}: NSGA3-CHS \citep{Cattelani2024BiB},
with Cox proportional hazards model as inner survival model (scikit-survival Python module implementation).
We used 5-fold CV to compare the considered MV strategies and view sets.
The MO algorithm aims to find all the best trade-offs of two objectives:
the C-index for the survival predictive ability,
and the root-leanness as a measure of parsimony in the number of features \citep{Cattelani2024BiB}.

\section{Supplementary results}

\mySupFig{sarc_baseline_best_comparison}{
	Expected versus measured C-index comparison of clinical-only and multi-omic feature selection strategies
    	in TCGA-SARC. Top panels show C-index as a function of the number of selected features.
    	“Expected” points represent internal cross-validation performance, while “measured” points denote testing
    	performance on the left-out sets. The results from all the 5 folds are plotted together.
    	Left column corresponds to clinical-only optimization, and right column
    	to the multi-omic sweeping strategy integrating clinical with expression-based features.
    	Bottom panels display the distribution of selected feature counts required to achieve increasing concordance levels,
    	illustrating model complexity and composition.
    	They show the counts and the expected C-index resulting from optimizing on the whole dataset.
}

\supFigRef{sarc_baseline_best_comparison} shows the comparison of the baseline with respect to the best
Sweeping* configuration according to the CHV metric in TCGA-SARC. The best CHV in this setting is obtained by
the concatenated configuration with clinic and mRNA views. Analogous comparisons for TCGA-KIRC and TCGA-LGG are
in Fig. 3.

The most frequently selected features for each Sweeping* configuration, after optimizing on the whole dataset,
are reported for TCGA-KIRC (\supFigRef{kirc_best_features}),
TCGA-LGG (\supFigRef{lgg_best_features}), and TCGA-SARC (\supFigRef{sarc_best_features}).

\mySupBigFig[1.0]{kirc_best_features}{
Top features identified for the TCGA‑KIRC cohort across all Sweeping* configurations and view combinations.
The figure lists, for each multi‑view (clinic; clinic + miRNA; clinic + miRNA + mRNA)
and Sweeping* configuration (concatenated; Sw; SwT; CSw; LCSw),
the 6 features most frequently selected in the approximated Pareto front
after optimizing on the whole dataset.
The more frequent the feature the darker the background.
}

\mySupBigFig[1.0]{lgg_best_features}{
Top features identified for the TCGA‑LGG cohort across all Sweeping* configurations and view combinations.
The figure lists, for each multi‑view (clinic; clinic + miRNA; clinic + miRNA + mRNA)
and Sweeping* configuration (concatenated; Sw; SwT; CSw; LCSw),
the 6 features most frequently selected in the approximated Pareto front
after optimizing on the whole dataset.
The more frequent the feature the darker the background.
}

\mySupBigFig[1.0]{sarc_best_features}{
Top features identified for the TCGA‑SARC cohort across all Sweeping* configurations and view combinations.
The figure lists, for each multi‑view (clinic; clinic + miRNA; clinic + miRNA + mRNA)
and Sweeping* configuration (concatenated; Sw; SwT; CSw; LCSw),
the 6 features most frequently selected in the approximated Pareto front
after optimizing on the whole dataset.
The more frequent the feature the darker the background.
}

\bibliography{reference}

\end{multicols}